\documentclass[preprint,5p,twocolumn]{elsarticle}
\usepackage{graphicx}
\usepackage{xcolor}
\usepackage{natbib}
\usepackage{color}

\usepackage{amssymb}
\usepackage{hyperref}
\usepackage{booktabs}
\journal{TBD}
\begin{document}
\title{Comparison of student performance between virtual and in-person modalities of introductory calculus-based physics I}
\author{Neel Haldolaarachchige}
\address{Department of Physical Science, Bergen Community College, Paramus, New Jersey 07652, USA}
\author{Kalani Hettiarachchilage}
\address{Department of Physics, Seton Hall University, South Orange, New Jersey 07079, USA}
\date{\today}

\begin{abstract}
Physics educators keep adding many skill developments to science and engineering students during their education as individuals and groups including critical thinking, conceptual understanding, problem-solving, mathematical implementation, computational implementation, \textit{etc}. Here, we are discussing how to reach and analyze students' outcomes within the context of introductory calculus-based physics courses by investigating two different teaching modalities. We found that there is no significant impact of teaching modality on student learning. By performing two different assessments: chapter-ending midterm assessments and unit-based (3-4 chapters) midterm assessments, shows that students can perform much better with short-time assessments in contrast to the long-time assessments. Further, we study any possible effects on students' final grades from students' prior knowledge of calculus and conceptual physics. This investigation shows that although there is no correlation between student's prior proficiency in calculus the class performance, however, there is a correlation of conceptual physics understanding towards class performance.
\end{abstract}


\maketitle
\section{Introduction}
Introductory level calculus-based physics classes are commonly offered and taught by traditional in-person modality and enhanced by adding different teaching methodologies towards students learning outcomes. Due to the COVID19 pandemic, all classes move to various online teaching formats such as online synchronous (meeting virtually on scheduled date and time), online asynchronous (no virtual meetings), hybrid (a portion of the class is done in-person and other portion done in virtual), and hybrid-rotational schedule (one group virtual and the other group in-person at the same time, then the two groups rotate in the following week).~\cite{liu, sim, faul} All of those modalities follow a well-structured schedule throughout the semester. Asynchronous classes allow students to learn on their own schedule, within a certain time-frame. Virtual synchronous classes allow students to learn from a distance, but they can virtually attend the class regularly, at the same time with their instructor and classmates. Synchronous classes have had strong attention in recent times mainly because such a modality can simulate the real in-person learning experience including class engagements and attractions between instructor-students, and between students. ~\cite{mar, ame, sara, yas}

In general, university-level introductory calculus-based physics class is one of the most challenging classes for most freshman (first-year student) and sophomore (second-year student) undergrad students where they gather many important concepts, skills, and training for their educations.~\cite{jess} Those students have to complete these classes as a requirement to advance into science and engineering majors. These physics classes consist of very important fundamental background knowledge to understand the development of scientific models and the physical processes of nature. Further,introductory-level physics classes are extremely important to build the analytical skills of future science and engineering students. Therefore, virtual class design and delivery methods have to be carefully structured and handled to achieve student learning outcomes. The use of learning management systems (LMS), which consists of functionalities to engage students in various learning activities, is an essential part of virtual teaching modality that can be used to elevate the student learning to the highest level by including various learning activities, group activities, easy access to course materials, easy access to running grades, instructor feed-backs, student engagements, \textit{etc}.\cite{talk1, talk2}

However, the main concern of the virtual teaching modalities lies in the assessments.~\cite{suta, hasan} There are various functionalities (lockdown browser, plagiarism checker, online proctoring software, \textit{etc.}) built into learning management systems (LMS) to support assessments and minimize or freeze possible academic integrity issues.~\cite{lms1, lms2, lms3, lms4, lms5, lms6} The problem is non of those plagiarism functionalities can be used in calculus-based physics assessments because students must solve very detailed problems that include drawing diagrams, building mathematical models, applying critical approaches, and finding algebraic solutions. In this paper, we address the importance of real-time proctoring implementation to overcome those problems.

Another important question is the correlation of students' background knowledge of basic physics concepts and prior proficiency in calculus towards the performance on the assessments.~\cite{bkp1, bkp2} This question becomes one of the heated topics when setting the prerequisites for university-level introductory calculus-based physics courses. Usually, at community colleges, only pre-calculus is a pre-requisite and calculus-I is a co-requisite for registering for the calculus-based physics I. There are no pre-requisites for conceptual physics or high school level but, it is recommended to complete the conceptual physics class prior to registration if students have no high school physics background. That means most students registered for calculus-based physics-I have no proper understating of the level of the class materials and expectations. Therefore, we investigate the correlation between students' prior knowledge of conceptual physics and their proficiency in calculus towards the academic achievement of calculus-based physics I (Mechanics) class.

This study was based on the introductory calculus-based physics-I at the largest community college, Bergen community college, in the state of New Jersey.  In general, calculus-based physics is considered the most challenging class for almost every engineering major student at Bergen community college. Historically, this class shows a very high dropout rate and failures. In this report, all engineering majors' achievements are observed, single instructor class evaluations were conducted but in different semesters, and similar sets of assessment, and proctoring were recycled.

This study focuses on three different research questions as given below. 
\begin{itemize}
    \item Does class modality (virtual or in-person) affect student performance? To investigate this comparison of student performance is conducted between in-person traditional and virtual-synchronous class settings.
    \item Does the type of assessment affect student performance? To investigate this, two types of mid-semester assessments were used (FA19 and FA20 – 3 mid exams, each 2.45 hrs) and SP21 (13 chapter exams, each 30 min). 
    \item How does a student's background knowledge of conceptual physics and proficiency in calculus 1 affect students' performance?
\end{itemize}

Here we present the details of the virtual class design and delivery methods of introductory calculus-based physics-I including the student performance and analysis. Then, detailed analysis with a comparison of two types of assessments is presented and discussed. Additionally, the correlation of class performance with students' background knowledge of conceptual physics and a prior proficiency in calculus are investigated.

\section{Methods}
\subsection{Use of learning management system}
Both in-person and virtual classes are fully web-enhanced with a learning management system (LMS).\cite{lms7} Among various LMS we use the popular two LMS; Moodle and Blackboard.~\cite{lms8, moodle, bb} All class activities including class materials and assignments were fully settled via LMS a way that students can navigate them easily. All assessments were submitted and graded electronically via LMS. The electronic pen was ~\cite{epen} used to provide feedback to students' submissions via LMS that simulate the real in-person paper grading and corrections. We discussed more details of the class design and delivery methods in the following references.~\cite{talk1, talk2}

\subsection{Real-time (live) proctoring of assessments via video conference technology}
Among available video conference technologies such as Zoom, Webex, Microsoft Teams, and Google Meeting.~\cite{vcd1}, we used the Webex platform to deliver the virtual synchronous classes.~\cite{vcd2} Most importantly class assessments (quizzes and exams) were proctored live via Webex.~\cite{lp1, lp2, lp3, lp4} Students were asked to frame the working area to the camera. The instructor can spot student behaviors during the quizzes/exams. During the quizzes/exams, students were asked to unmute audio and not to use headphones to confirm that they do not discuss the problems with someone else. They are not allowed to use any notes other than the provided equation sheet for the exams.

\subsection{Teaching methods and learning support}
The study was based on the physics I (Mechanics) class conducted in three semesters (FA19, FA20, and SP21). Fall 2019 was an in-person class and both fall 2020 and spring 2021 were virtual synchronous classes. Classes were taught by the same instructor and followed the same teaching style. The teaching style used for the class is: start with a brief review of previous class materials, then talked over the plans for the day with emphasizing important concepts briefly, next discussed selected concept/principle/theory in detail followed by related problem-solving as a discussion. Each step of the problem-solving process is conducted by requesting students feedback, group work, and participation. The instructor randomly went around the class to check students' work and help out with students' difficulties.

Computer simulations were used during the short lecture to visualize most of the concepts. More opportunities to simulate, collect and analyze data were given to the students during the laboratory period. It is found that initializing concepts through computer simulations is the best effective way to start a new topic.

Since the introductory physics classes are combined with the set of aligned experiments, special emphasis was given to the development of teaching resources for experiments. To achieve this goal new series of simulation-based lab manuals were developed and a series of lesson videos are provided with a step-by-step process of simulation, data collection, and data analysis with Microsoft Excel.~\cite{sbe1, sbe2, sbe3}

Students' learning was supported with a few different methods. Extended office hours were conducted by the instructor throughout the semester to help students with difficulties in understanding concepts and problem-solving. Further,  students were given opportunities to attend supplemental instruction (SI) sessions which are similar to recitation in most of the colleges. Usually, there were two 2 hour SI sessions per week. The sessions were conducted by a senior undergraduate student or a professional tutor who successfully completed the same course with a letter grade of B+ or A. Additionally, students were given extra support at the college tutoring center during regular hours with physics tutors.~\cite{si1, si2}

\subsection{Data analysis and surveys}
Microsoft Excel software package was used for data analysis and all the tables and graphs were completed with Excel. A survey at the beginning of the semester was used to collect student responses to understand their background knowledge of conceptual physics and calculus. The survey was posted for students within the first week of the semester via LMS and responses were collected anonymously. Students' response rate is 100\%. The results were investigated to find a correlation of students' background knowledge with class performance.

\subsection{Type of assessments and performance analysis}
The student performance is investigated by using the results of the virtual-synchronous introductory physics I class from two consecutive semesters (FA20 and SP21, both during COVID19). Various types of assessments as shown in table~\ref{tab:Grade} were used to analyze student performance. Next, the results are compared to the same in-person traditional class in FA19 (before COVID19). Analysis of assessments is conducted in two aspects to understand the student performance, 1) percent of students below and above 70\% of the mid-semester assessments and 2) percent of students below and above 70\% of the cumulative final exam. Mid-semester assessments were done with a similar set of questions in all three semesters. However, the method of mid-semester assessments was changed in SP21 as follows, during FA19 (in-person) and FA20 (virtual-synchronous) there were three 2.45 hour-long timed closed-book mid-semester assessments, and during SP21 there were thirteen 30 minute-long timed closed-book chapter-based mid-semester assessments. Additionally, in SP21 semester,  optional practice questions for each chapter were provided to the students and more than 90\% of students completed those weekly.

\begin{table}[!htp]
\caption{\label{tab:Grade}The final class grade equation}
\begin{tabular*}{\columnwidth}{@{\extracolsep{\fill}}lc}
\toprule
  Description& score(\%)\\
\hline
Mid term assessment & 40 \\ 
(SP21 – 13 in-class quizzes and \\
FA19 and FA20 – 3 Mid exams) \\
Final Exam (cumulative)$\:$& 25\\
Homework& 5\\
Pre-lecture reading quizzes& 5\\
Labs~(datasheets, and detail reports)& 25\\
Total& 100\\
\hline
\end{tabular*}
\end{table}


The final exam format (3.00hrs long, timed, and closed-book) was the same for all three semesters. New exam questions (for both mid-exams and final exams) were written from scratch for all three semesters but the format, concepts, and theory/principles were kept the same. Each exam question was inspected in advance via Google to ensure that nothing popups with an exact search. The written solutions were uploaded to LMS by students and carefully analyzed by the instructor. No academic integrity issues were detected. All the mid-exams (or chapter exams) and final exams were proctored live via video conference. In a virtual class setting, the use of a new set of questions and live proctoring provide a quality education for all students and that minimizes or eliminates any possible academic integrity concerns.
 
Each student's mid-exams (or chapter exams) total and final exam total were normalized relative to a passing score of 70\%. This makes it easier to find the total number of students who got above and below the passing score (70\%) because the normalized value is positive for students with above the passing score (70\%) and it is negative for students with below-passing score (70\%). Then, the percentage of students who got above and below the passing score (70\%) was calculated.

\section{Results and Discussion}
\subsection{Analysis of student performance in mid-semester assessments and final exam}
One of the most interesting and important questions is the effect of class modality (in-person and virtual-synchronous) on student learning outcomes. We investigated this question by comparing three semesters of class averages of mid-semester and final assessments between in-person vs virtual-synchronous class modalities.

\begin{figure}[!htp]
  \centerline{\includegraphics[width=0.5\textwidth]{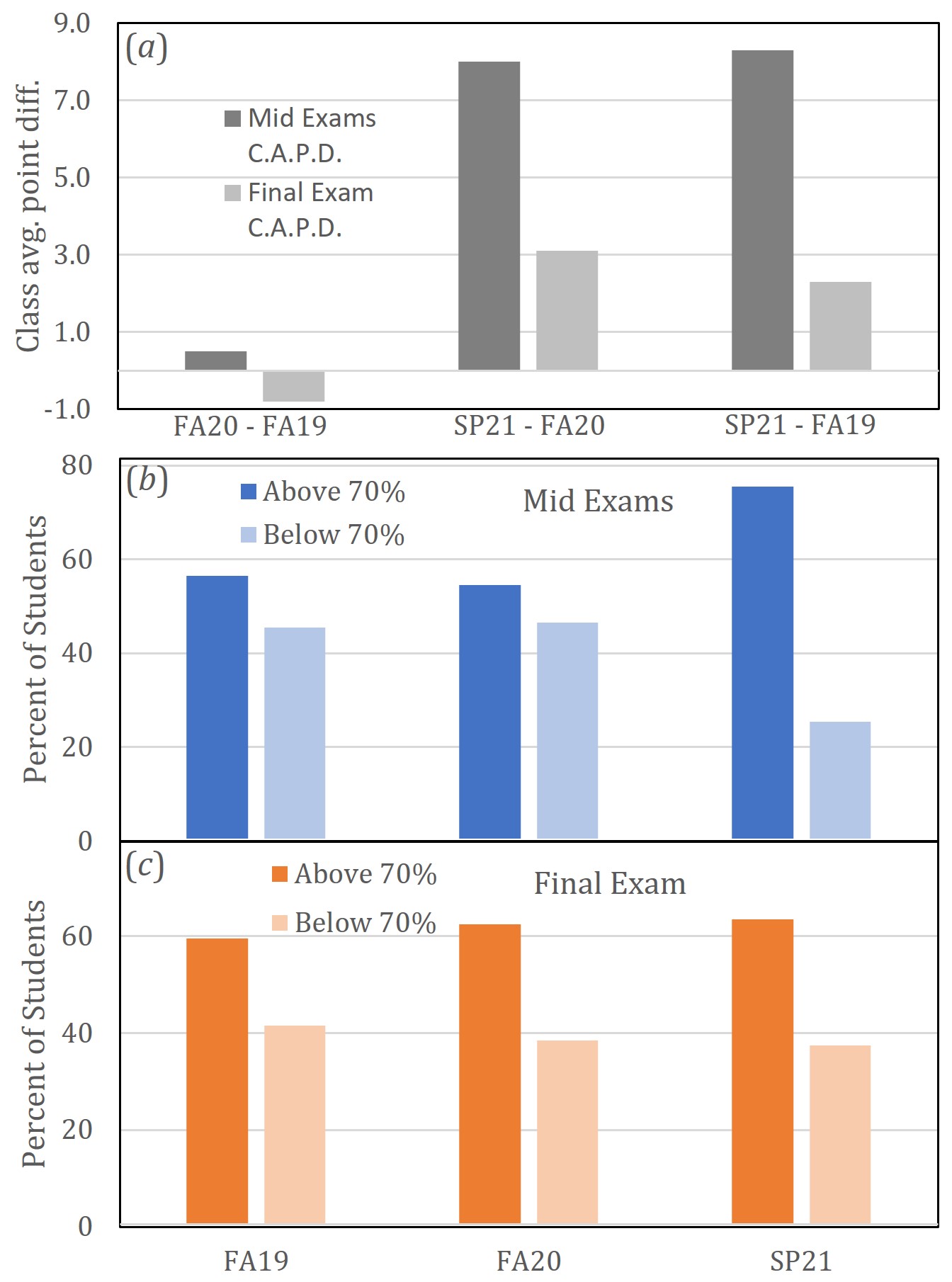}}
  \caption
    {
    (Color online) (\textit{a}) The class average percent difference (CAPD) between the semesters. The dark gray color shows CAPD of mid exams and light gray color shows CAPD of the final exam. (\textit{b}) Percentage of students who scored above 70\% (bright blue color) and below 70\% (light blue color) of mid-exams. (\textit{c}) Percentage of students who scored above 70\% (bright orange color) and below 70\% (light orange color) of final exams.
   }
\label{fig1}
\end{figure}

Figure~\ref{fig1}(\textit{a}) shows an analysis of class average percent difference (CAPD) of mid-exams (dark gray color) and the final exam (light gray color) between semesters. It can be observed that CAPD of mid-exams and final exams between FA19 (in-person) and FA20 (virtual) is negligible. On the other hand, CAPD is very high between SP21 (virtual) vs FA20 (virtual) and SP21 (virtual) vs FA19 (in-person). CAPD is significantly high in mid-exams and considerably improves in final exams. This shows that something very specific happens during SP21 but, it does not correlate with the type of class modality since the CAPD of final exams is the same between virtual two semesters (FA20 and SP21) and virtual-in-person two semesters (SP21 and FA19).

Analysis of student performance in three semesters (FA19-in-person, FA20, and SP21 - virtual synchronous) of mid-exams shown in figure~\ref{fig1}\textit{b}. The percentage of students with above 70\% of mid-exams (the bright blue color) is about the same in FA20 (virtual) and FA19 (in-person). This shows that the class type (virtual or in-person) does not affect student learning. However, students' percentage above 70\% in mid-exams is significantly higher in SP21 compared to that of FA20 and FA19. This indicates that SP21 semester results show significant improvement in student performance. The increment of the passing percentage of students in SP21 correlates with the format of mid-exams. That means chapter-based mid-exams (short-time and close-book) enhance student permanence. Both the other semesters (FA19 and FA20) mid-exams were unit-based exams (3-4 chapters per mid-exam, long-time, and close-book). Therefore, we suggest that the chapter-based mid-exams are more effective than the unit-based exams towards student achievement in calculus-based physics classes.

On the other hand, the final exam analysis of three semesters (FA19-in-person, FA20, and SP21 - virtual synchronous) shows in figure~\ref{fig1}(\textit{c}). It can be observed that the percentage of students above 70\% score is very similar in all three semesters. This is a very good indication that the class modality has no effect on student learning. Further, this shows that students do not perform well in longer-timed and closed-book exams (final exams are 3.0 hrs long and cumulative). When comparing the percentage of students above 70\% score of mid-exams and final exams in the SP21 semester, we observe a significant difference. The higher percentage of students above 70\% score in mid-exams during the SP21 semester shows that students can perform well in short-time close-book exams.

\subsection{Analysis of student performance with final class grades}
\begin{figure}[!htbp]
  \centerline{\includegraphics[width=0.5\textwidth]{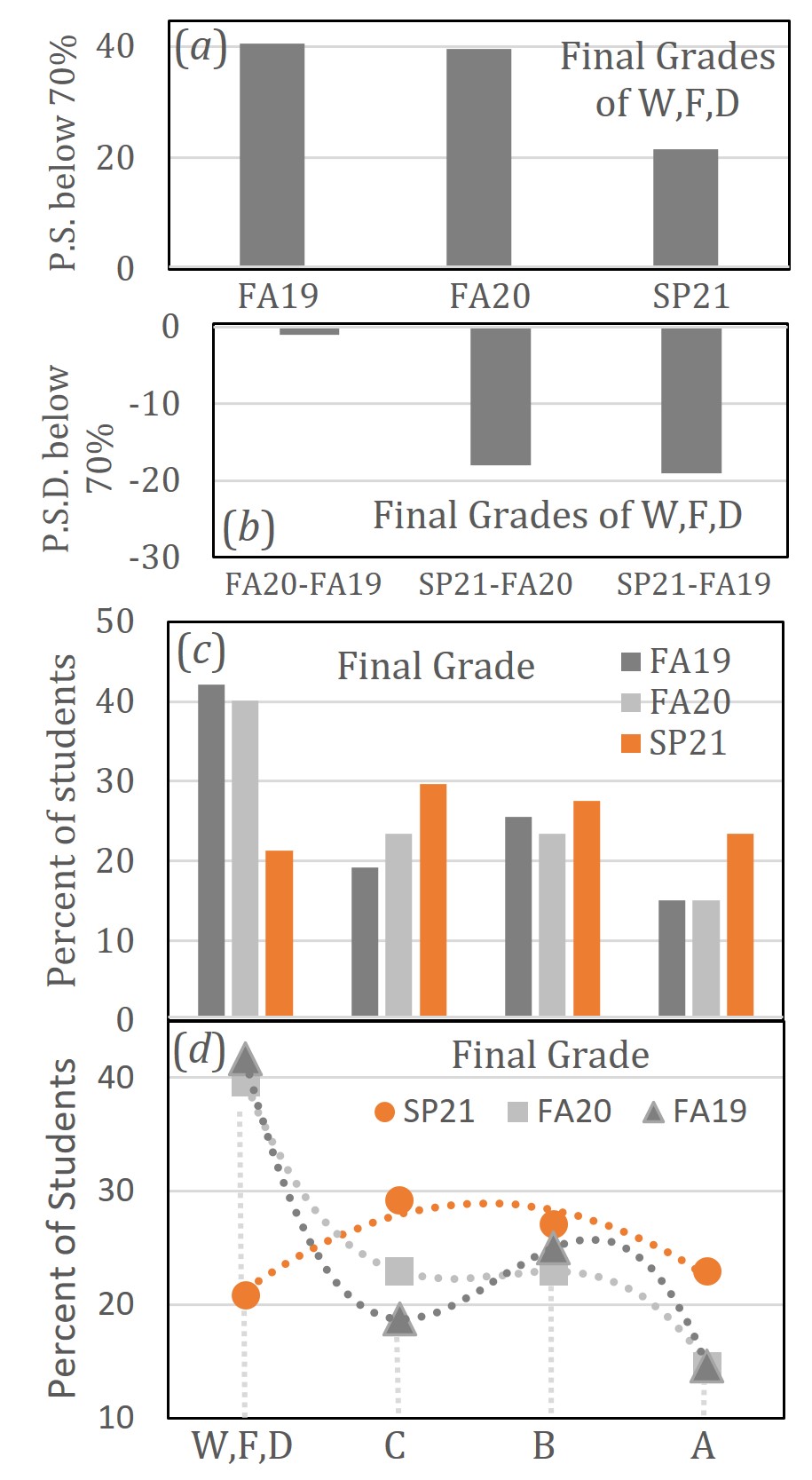}}
  \caption
    {
    (Color online) Final grade analysis of three semesters (FA19-in-person, FA20, and SP21 - virtual synchronous). (\textit{a}) shows the percentage of students (P.S) below 70\% at the end of the semester. (\textit{b}) shows the percentage of students difference (P.S.D) between two semesters below 70\% at the end of the semester. (\textit{c}) displays the final grade as a function of students' percentage for all three semesters (FA19 in dark gray, FA20 in light gray, and SP21 in orange). (\textit{d}) shows scatter plot of students percentage as a function of the final grade for all three semesters (FA19 in dark gray, FA20 in light gray, and SP21 in orange and dotted line represents the polynomial fitting. Letter grades W = withdrawal, F = below 70\%, and D = between 60\% - 70\%).
   }
\label{fig2}
\end{figure}

Figure~\ref{fig2}(\textit{a} and \textit{b}) show the analysis based on end of the semester class grades: the percentage of students (P.S.) who received below 70\% (including grades of W - withdrawal, F - less than 60\%, and D - between 60\% and 70\%). Figure~\ref{fig2}(\textit{a}) shows total percentage of students below 70\% of the score. SP21 (virtual-synchronous) semester shows better student performance compared to the previous two semesters (FA19 - in-person and FA20 - virtual synchronous). The percentage of students' difference (P.S.D) with lower than 70\% between two semesters can be seen in figure~\ref{fig2}(\textit{b}). It shows that the failure percentage has been dropped by 20\% in SP21 when compares to the previous two semesters (FA19 and FA20) but, there is no difference between FA19 (in-person) and FA20 (virtual-synchronous). It shows that the student's performance enhances significantly in the SP21 semester. As shown with the analysis in figure~\ref{fig1}, the reason for better student performance in the SP21 is the type of mid-semester assessments.

Figure~\ref{fig2}(\textit{c} and \textit{d}) show the final class grade analysis for all three semesters (FA19 - dark gray color, FA20 - light gray color, and SP21 - orange color). The final class grade was calculated according to the equation given in Table~\ref{tab:Grade}. The final grade equation is same for all three semesters except during SP21 mid-semester assessment was based on chapter quizzes. There is no significant difference in final class grade distribution between FA19 (in-person) and FA20 (virtual) semesters. Both graphs show that the percentage of students below 70\% (letter grades of W, F, D) is very similar in FA19 (in-person, pre-COVID) and FA20 (virtual-synchronous, during COVID). Further, it shows that there is no possible bell curve (see figure~\ref{fig2}\textit{d}) for the distribution of final grades in FA19 and FA20. This is due to a large percentage (about 40\%) of students below 70\%. However, grade distribution in SP21 (virtual-synchronous) shows a nice bell curve as in figure~\ref{fig2}(\textit{d}). The student percentage below 70\% with letter grades W, F, and D in SP21 is dropped by about 20\% compared to FA20 and FA19. This is a significant improvement in student performance. This improvement in SP21 should be a effect of mid-semester assessment (chapter quizzes) as shown in figure~\ref{fig1}(\textit{b}) because final exam itself shows no significant difference in all three semesters (see figure~\ref{fig1}(\textit{c}).

\subsection{Effect of type of assessment to student performance}
The analysis of mid-semester assessments and final exams in figure~\ref{fig1} and final class grades in figure~\ref{fig2} shows that there is a significant improvement in students' performance in the SP21 semester compared to the others. Here, we discuss the reasons for this achievement. The Classes were conducted by the same instructor with the same teaching style in all three semesters. The same resources were used for teaching for all three semesters. The final exam assessments (3.0 hrs, proctored, close-book) were the same in three semesters. Similar student performance in final exams (see figure~\ref{fig1}\textit{c}) in all three semesters showed that there were no academic integrity concerns even though the final exams were conducted virtually during FA20 and SP21. This also shows that live proctoring exams via video conference technology is an effective way to conduct the exam in virtual classes.

Now, let's take a closer look at the comparison of student performance between two types of assessments. In general, there are three mid-exams (each exam 2.45 hrs, closed-book, proctored, and covers 3-4 chapter content) in calculus-based physics classes, and from here on called long-time assessments. That type of mid-exam was used during FA19 (in-person), and FA20 (virtual-synchronous). However, the type of mid-semester assessments was changed to chapter exams during SP21 (virtual-synchronous) which were short-time (each 30 min), closed-book and proctored, and from here on called short-time assessments. Additionally, at the end of the semester SP21, the two lowest chapter quizzes were dropped. The effect of the short-time continuous assessments throughout the semester can be observed in figure~\ref{fig1}(\textit{b}). The continuous mid-semester assessments affect most of the students to perform better at the end of the semester which can be observed in figure~\ref{fig2}. Also, figure~\ref{fig2}(\textit{d}) shows the bell curve nature of final grade distribution in the SP21 semester, which did not display in the other two semesters (FA19 and FA20). Therefore, we can assume that the enhancement of student performance correlates with continuous assessment in the SP21. On the other hand, our comparison results of all three semesters show that there is no impact of class modality on student performance.

\subsection{Analysis of students' background knowledge of conceptual physics and calculus on class performance}
Another important and interesting aspect of this study is the effect of student background knowledge of conceptual physics and calculus on the performance of university-level calculus-based physics. The research question that we addressed here is as follows, “How does a student’s background knowledge of conceptual physics and proficiency in calculus 1 affect students’ performance?”. The survey questions on background knowledge of conceptual physics and calculus were done in the very first week of the semester. Students who did not complete calculus I at the high school level were simultaneously taking this class and calculus I at the college/university level. Survey did not include a question on background asking students if they have done both Physics and calculus at the high school level. 

\begin{figure}[!htbp]
  \centerline{\includegraphics[width=0.5\textwidth]{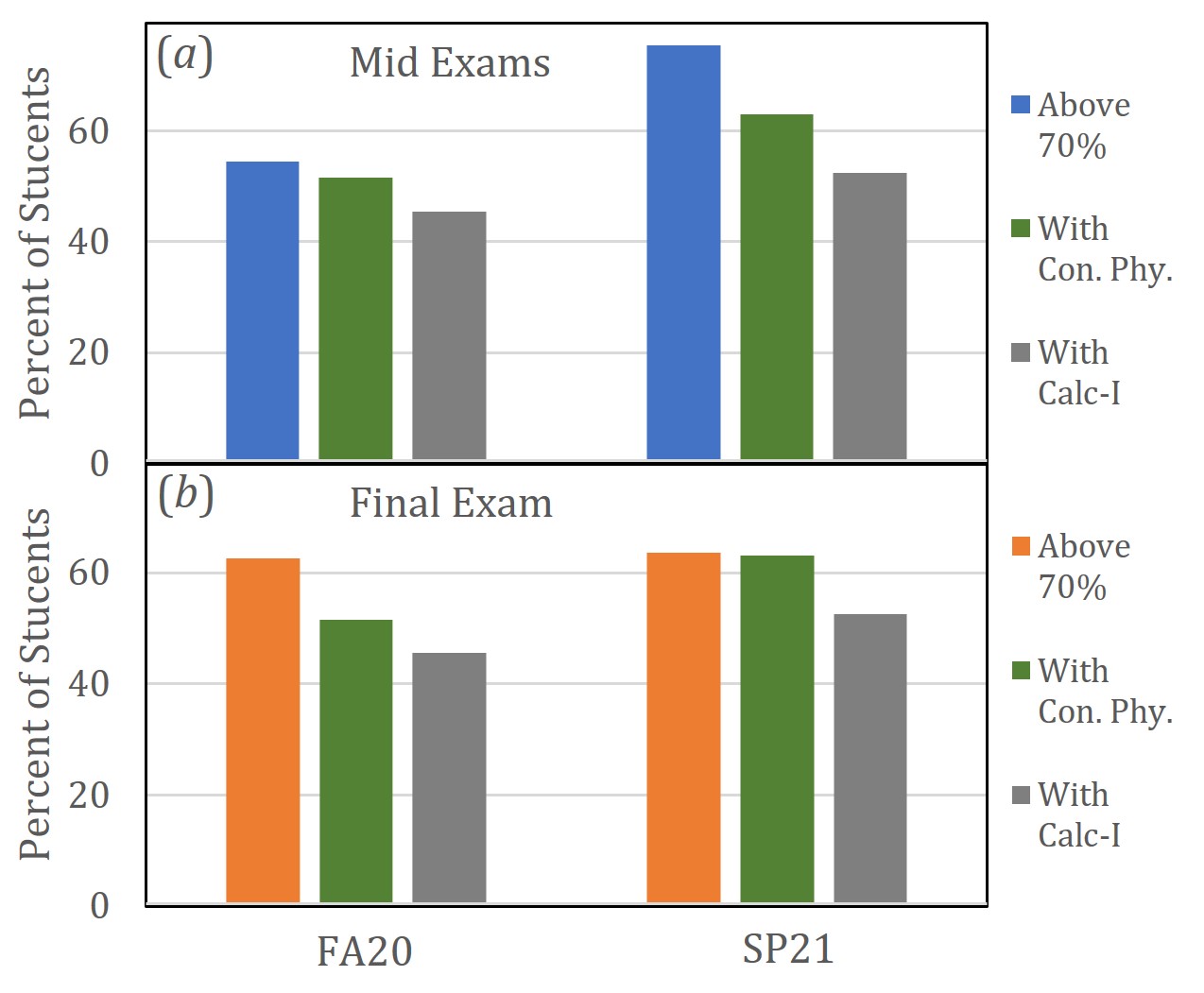}}
  \caption
    {
    (Color online) (\textit{a}) Student percentage above 70\% in mid-exams (blue color) compared to student background knowledge of conceptual physics (green color) and calculus-I (gray color). (\textit{b}) Student percentage above 70\% in the final exam (orange color) compared to student background knowledge of conceptual physics (green color) and calculus-I (gray color).
    }
\label{fig3}
\end{figure}

Figure~\ref{fig3}(\textit{a}) shows student performance in mid-semester assessment with the percent of students who completed conceptual physics (green color) or calculus I (gray color) before starting with a calculus-based physics class. It shows that the passing parentage of students is higher than both percentages of students with conceptual physics and calculus.
When observing SP21 students passing percentage in mid exams it can be seen that about 8\% of students pass the class with out conceptual physics background and more than 16\% of students pass the class with out calculus I level math knowledge. Based on the above mentioned reason we suggest that concept-physics background correlates well with class performance than that of the calculus background. 

Students with conceptual physics background in SP21 (see figure~\ref{fig3}\textit{a}) are higher than the students with calculus-I level math proficiency in both FA20 and SP21. This can be an effect of three reasons, 1) the background of incoming freshmen population changes every year, 2) in SP21 there were more students in the class at the end of the semester than in the FA20 (low dropout rate), and 3) calculus-I is only a co-requisite for calculus-based physics-I, most students simultaneously taking both calculus-based physics I and calculus-I. Also, figure~\ref{fig3}\textit{a} shows that student passing percentage on mid-exams is much higher than calculus proficiency but only slightly higher than conceptual physics knowledge. As discussed earlier, the reason for this is also the impact of student performance from short-time mid-semester assessments in SP21 compared to the long-time mid-assessments in FA20. It shows that students who started with a background in conceptual physics perform well in mid-exams than that the students who started the class with the background in calculus-I level math.

 The percentage of students with passing grades on the final exam in FA20 (see the left side of figure~\ref{fig3}\textit{b}) is higher than the student percentage with background knowledge of conceptual physics and calculus-I. However, figure~\ref{fig3}(\textit{b}) (see right side) shows that the student percentage with passing grades on the final exam is very similar to the percentage of students with conceptual physics backgrounds in the SP21 semester. Also, it shows that the percentage of students with calculus-I is much lower than the passing percentage of the students in the final exam in the SP21 semester. This particular observation suggests two important points; 1) conceptual physics knowledge is more relevant to better in calculus-based physics, and 2) further studies are needed to better understand why some students can do well in calculus-based physics without prior knowledge of both conceptual physics and calculus. However, analysis in figure~\ref{fig3} indicates that student performance in the final exam correlates well with the conceptual physics background than with calculus-I knowledge.

\section{Conclusion}
Analysis of student performance in mid-semester assessments and the final exams were completed and compared in three semesters. The results suggest that there is no effect of class modality (in-person or virtual-synchronous) on student performance. This implies that calculus-based physics classes can be conducted virtually as effectively as in an in-person setting. This is a very good conclusion for community colleges (or two-year colleges) since 
almost all students in community colleges are working either full or part-time and also have other family commitments due to living with families and kids. Therefore, time commitment to college and studying for those students is a significant problem. That can be easily addressed by offering the classes in synchronous format at the same time providing them the best quality education which can simulate the same learning experience as on-campus classes.

Further, the results suggest that online live proctoring of assessments with video conference technology can be used to minimize (or eliminate) academic integrity concerns. That is one of the primary problems and concerns with virtual classes of technical subjects. Standard software-based proctoring methods do not work for virtual calculus-based physics, mainly because assessments are based on open-ended detail problem-solving.

Additionally, the results show that short time chapter-based mid-exams are a more effective way to reach expected student outcomes than the traditional long-standing cumulative mid-semester assessment method. In this study, we compared students' performance between cumulative (3-4 chapter contents) and non-cumulative (single chapter content) mid-semester assessments to conclude that the chapter-based method is more effective. This result is very important for college educators because most college students are facing a hard time in their first calculus-based physics class and the failure rate of such classes is historically high. That problem could be solved by changing the methodology of assessments as discussed above. Further, it shows that students can absorb materials better when they prepare for short chapter-based exams. This study shows that a higher percentage of students complete the class with passing final grades if the mid-semester assessments are based on short-time (non-cumulative) chapter exams.

Another aspect that we tested here is a correlation of students' performance with prior knowledge of physics and calculus. The results suggest that the students who started the class with a conceptual physics background can perform better in class assessments than the students who started without a conceptual physics background. On the other hand, the results also suggest that there is no direct correlation between student performance with calculus-I proficiency. This is an important result because it suggests that it may be worth revisiting the pre-requisites of first calculus-based physics at the college level. Also, this outcome can be used to create more effective learning support mechanisms for students. Dedicated supplemental instruction or recitation for calculus-based physics seems to be very important and could play a major role in enhancing student learning.

\section{Acknowledgments}
NH acknowledges the Cerullo Learning Assistance Center (CLAC) that provided learning supports (supplemental instruction - SI and tutoring) to students. Supplemental instruction (SI) was financially supported by the STEMatics grant from the Department of Education.   
NH acknowledges the academic freedom and support provided to investigate new teaching methods by the dean of the STEM division Dr. Emily Vandalovsky (and former dean - Dr. P.J. Ricato) and department chair of Physical science Dr. Ara Kahyaoglu (and former chair - Dr. Lynda Box) at Bergen community college. 

\section{Ethical Statement}
This research was done according to ethical principles outlined by Institutional Review Board (IRB) at Bergen Community College. That is in accordance with the ethical policies outlined by the Institute of Physics (IOP). This article does not contain any personal data, and it does not identify any individuals (students and/or teachers). All students and teachers have given explicit consent to participate in a research project.



\begin{thebibliography}{100}
\bibitem{epen} A pen like electronic item can be used write on a computer screen. Electronic pen is also called stylus in which tip position on the computer screen can be detected and accordingly computer can identify what is written on the screen.
\end{thebibliography}
\end{document}